\documentclass[lettersize,journal]{IEEEtran}
\usepackage{amsmath,amsfonts}
\usepackage{algorithmic}
\usepackage{algorithm}
\usepackage{array}
\usepackage[caption=false,font=normalsize,labelfont=sf,textfont=sf]{subfig}
\usepackage{textcomp}
\usepackage{stfloats}
\usepackage{url}
\usepackage{verbatim}
\usepackage{graphicx}
\usepackage{cite}
\usepackage{lineno}
\usepackage{xcolor}
\begin{document}

\title{Development of an MKID frequency-to-pixel LED mapper for SPT-3G+}

\author{E. S. Martsen, P. S. Barry, B. A. Benson, K. R. Dibert, K. N. Fichman, T. Natoli, M. Rouble, C. Yu \\
\textit{On Behalf of the SPT-3G+ Collaboration}
\thanks{(Corresponding Author: Emily Martsen))}
\thanks{E. S. Martsen, K. R. Dibert, K. N. Fichman, and T. Natoli are with the University of Chicago, 5640 South Ellis Avenue, IL,
60637, USA and Kavli Institute for Cosmological Physics, U. Chicago,
5640 South Ellis Avenue, Chicago, IL, 60637, USA (email: emartsen@uchicago.edu, krdibert@uchicago.edu, kfichman@uchicago.edu, tnatoli@uchicago.edu)}
\thanks{P. S. Barry is with Cardiff University, Cardiff CF10 3AT, UK (email:
barryp2@cardiff.ac.uk)}
\thanks{B. A. Benson is with Fermi National
Accelerator Laboratory, PO BOX 500, Batavia, IL 60510, the University of
Chicago, Chicago, IL, 60637, USA, and Kavli Institute for Cosmological
Physics, U. Chicago, 5640 South Ellis Avenue, Chicago, IL, 60637, USA
(email: bbenson@astro.uchicago.edu,)}
\thanks{C. Yu is with the Argonne National Laboratory, Argonne,
IL 60439 USA, the University of Chicago, 5640 South Ellis Avenue,
Chicago, IL, 60637, USA, and Kavli Institute for Cosmological Physics,
U. Chicago, 5640 South Ellis Avenue, Chicago, IL, 60637, USA (email: cyndia.yu@anl.gov)}
\thanks{M. Rouble is with McGill University, 845 Rue Sherbrooke O, Montreal,
QC H3A 0G4, Canada (email: maclean.rouble@mcgillcosmology.ca)}
}

\markboth{Journal of \LaTeX\ Class Files,~Vol.~14, No.~8, August~2021}%
{Shell \MakeLowercase{\textit{et al.}}: A Sample Article Using IEEEtran.cls for IEEE Journals}


\maketitle

\begin{abstract}
SPT-3G+ is the next-generation camera for the South Pole Telescope (SPT). SPT is designed to measure the cosmic microwave background (CMB) and the mm/sub-mm sky. The planned focal plane consists of 34\thinspace000 microwave kinetic inductance detectors (MKIDs), divided among three observing bands centered at 220, 285, and 345 GHz. Each readout line is designed to measure 800 MKIDs over a 500 MHz bandwidth, which places stringent constraints on the accuracy of the frequency placement required to limit resonator collisions that reduce the overall detector yield. To meet this constraint, we are developing a two-step process that first optically maps the resonance to a physical pixel location, and then next trims the interdigitated capacitor (IDC) to adjust the resonator frequency. We present a cryogenic LED apparatus operable at 300 mK for the optical illumination of SPT-3G+ detector arrays. We demonstrate integration of the LED controls with the GHz readout electronics (RF-ICE) to take data on an array of prototype SPT-3G+ detectors. We show that this technique is useful for characterizing defects in the resonator frequency across the detector array and will allow for improvements in the detector yield. 
\end{abstract}

\begin{IEEEkeywords}
Superconducting microwave devices, Other non-equilibrium (non-thermal) detectors (e.g. SIS, MKID), Instrumentation and readout of superconducting devices, Multiplexing
\end{IEEEkeywords}

\section{Introduction}
\IEEEPARstart{S}{PT-3G+} is the next-generation camera to be put on the South Pole Telescope (SPT), a 10-meter telescope located at the South Pole \cite{anderson_spt-3g_2022, benson_spt-3g_2014}. SPT-3G+ will complement existing SPT data by making measurements at 220, 285, and 345 GHz, beyond the peak of the CMB blackbody spectrum, which will enable particularly powerful constraints on secondary CMB anisotropies, Galactic dust, and the cosmic infrared background \cite{anderson_spt-3g_2022}. The planned focal plane will consist of $34\thinspace000$ microwave kinetic inductance detectors (MKIDs) across seven wafers composed of six triangular sub-arrays\cite{anderson_spt-3g_2022}. This instrument design has 800 detectors per 500 MHz of readout bandwidth, which requires precise control of detector resonance frequency.

\begin{figure}
    \centering
    \includegraphics[width=0.7\linewidth]{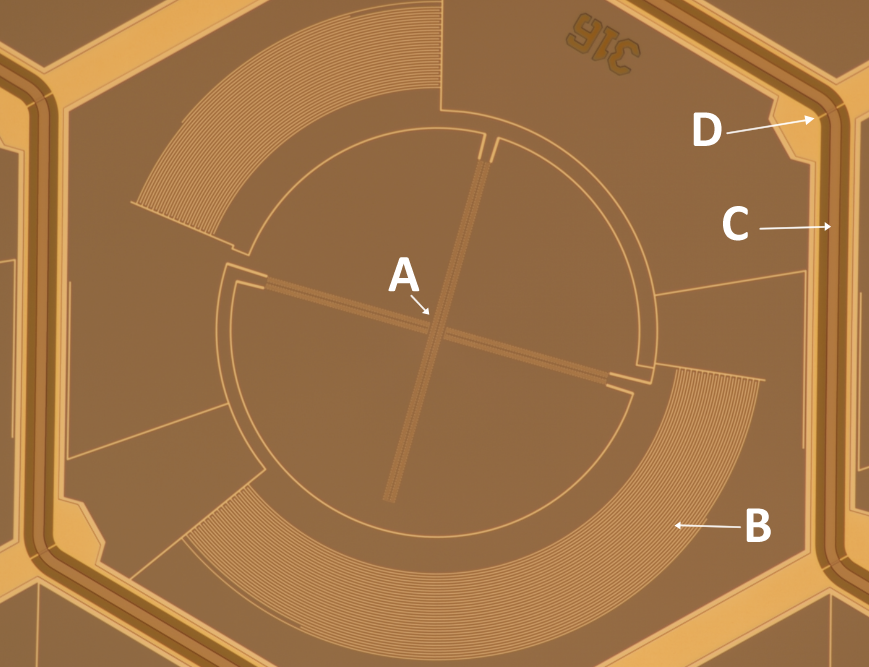}
    \caption{Microscope image of an SPT-3G+ MKID 220 GHz pixel. The pixel contains two orthogonally aligned inductors (A) and two arc-shaped IDCs (B), which are in turn coupled to the coplanar waveguide (CPW) feedline (C) which is bridged at the vertices of the hexagonal pixel cell (D) to maintain ground-plane connectivity \cite{dibert_characterization_2023}.}
    \label{fig:mkid_microscope}
\end{figure}
The SPT-3G+ pixel design consists of two lumped element MKIDs with aluminum orthogonal polarization absorbers and interdigitated capacitors (IDCs) coupled to a niobium coplanar waveguide feedline (Fig. \ref{fig:mkid_microscope})\cite{dibert_characterization_2023,dibert_development_2022}. 

\begin{figure}
        \centering
        \includegraphics[width=0.9\linewidth]{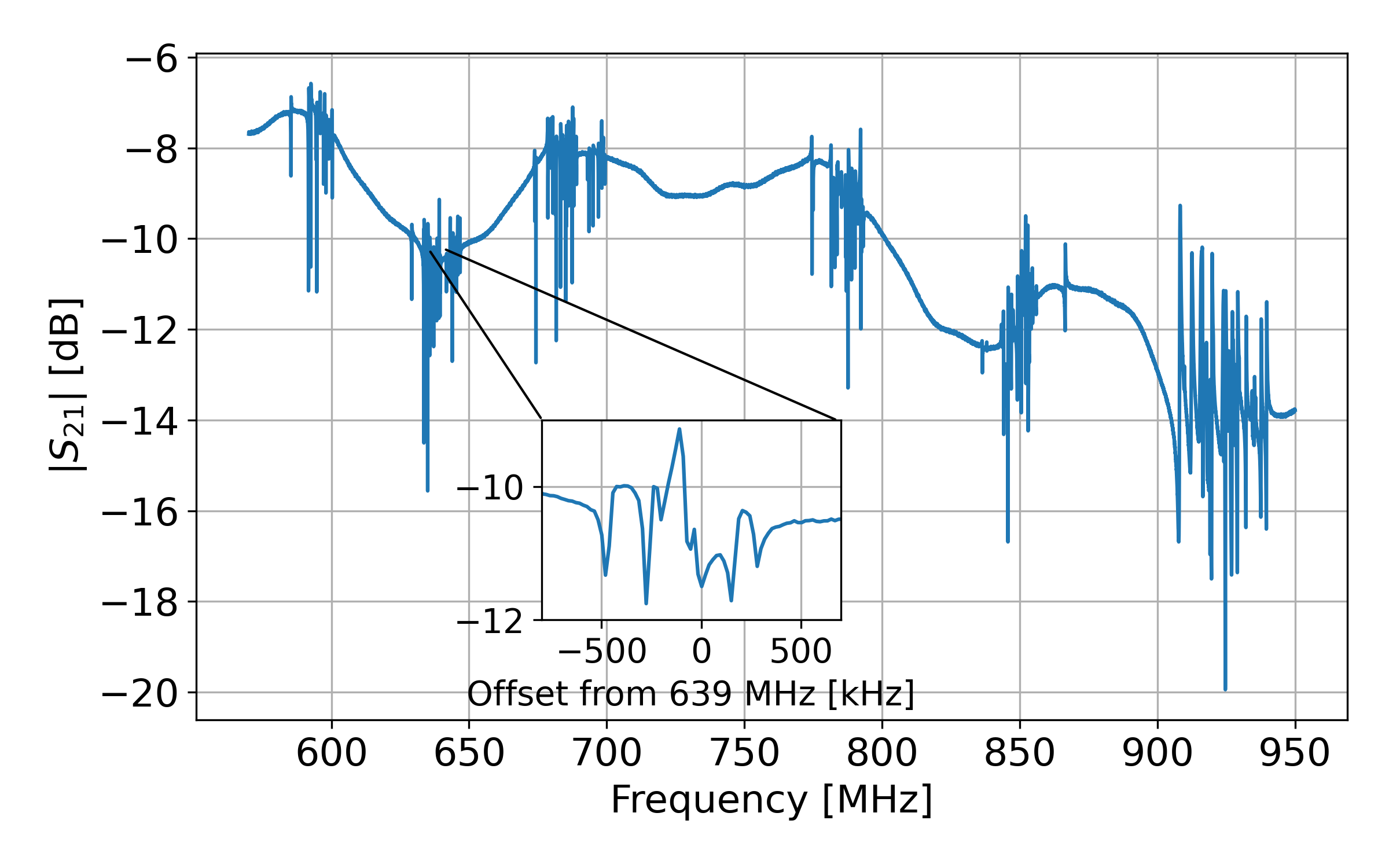}
        \caption{Frequency sweep from SPT-3G+ device. Frequencies are split into 6 visible banks to separate resonances occupying the same unit cell. The distance between frequency banks is exaggerated for the purpose of this demonstration. Inset shows clashing resonators.}
        \label{fig:netanal}
    \end{figure}

By coupling each MKID in a subarray to the same feedline at a unique resonance frequency, they can all be readout through a single radio frequency (RF) channel (Fig. \ref{fig:netanal}), allowing for multiplexing in the frequency domain. However, a requirement to meet the multiplexing goals of SPT-3G+ is the precision of the fabricated resonator frequencies, which becomes challenging at the 566 kHz spacing required given our RF readout and multiplexing factor. The main challenge in acquiring 566 kHz spacing comes from resonance scatter.   
Resonance scatter is the characteristic resonance frequency of an MKID shifts from its intended position. The cause of such scatter is generally believed to stem from inconsistencies during the fabrication process, but the exact mechanism is often difficult to pinpoint \cite{Liu2017,mcgeehan2018}. When scatter is not controlled, the resonances swap orders or clash into each other, rendering detectors unusable (Fig. \ref{fig:netanal} inset). 
One method that has been implemented to control resonance scatter is through the optical identification of physical pixels to frequency-space resonator channels using LEDs and subsequent post-processing to rearrange the frequency ordering to match the intended design \cite{liu_cryogenic_2017, Liu2017}. We present an LED mapper design for SPT-3G+ and demonstrate mapping SPT-3G+ detector arrays at 300 mK, a higher operating temperature than previous MKID LED mapping studies have used. We  additionally present a method for mapping multiple pixels with each LED to reduce the number of LEDs needed.

\section{Methods}
\subsection{Mapper Design}

\begin{figure}
        \centering
        \includegraphics[width=0.7\linewidth]{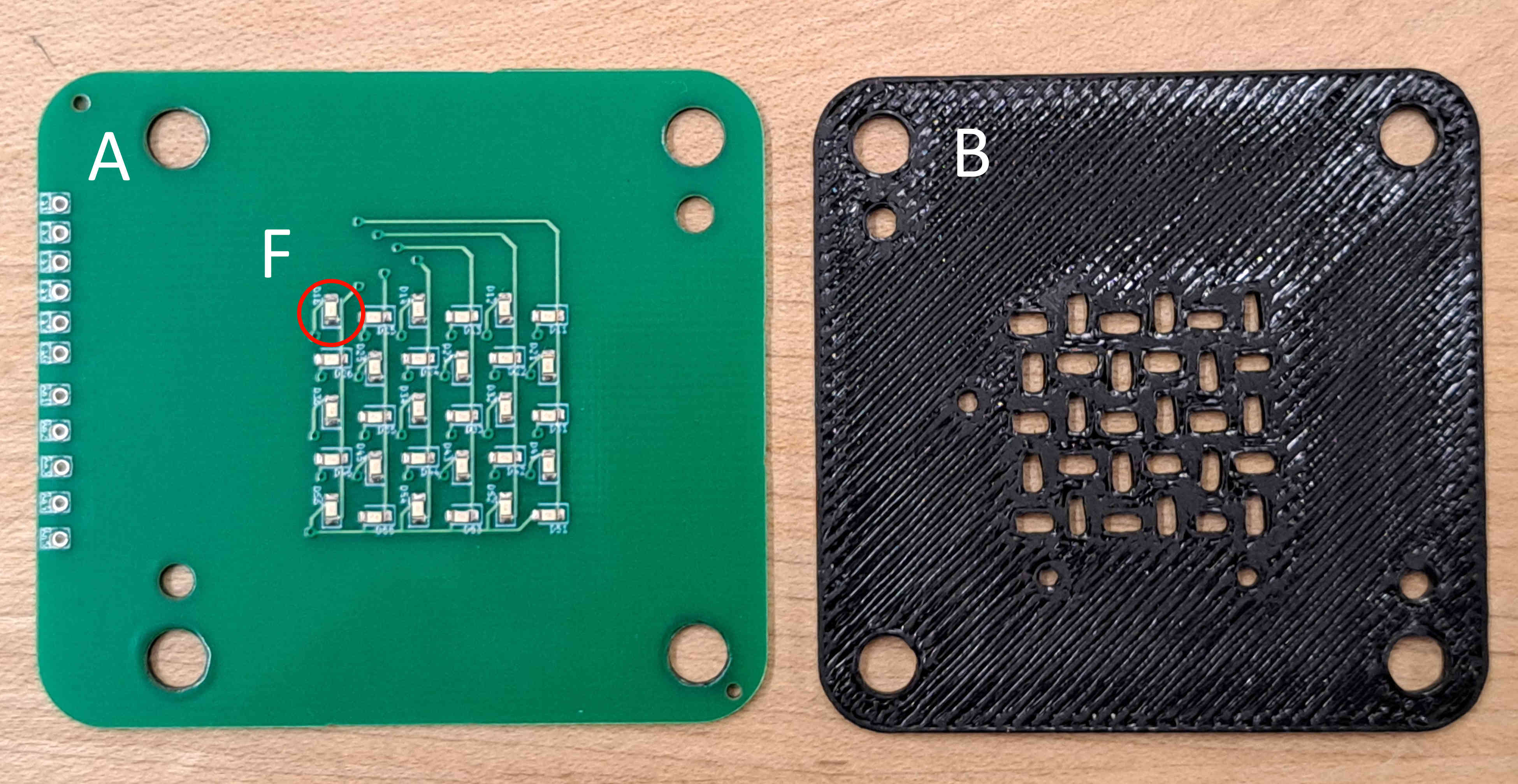}
        \includegraphics[width=0.8\linewidth]{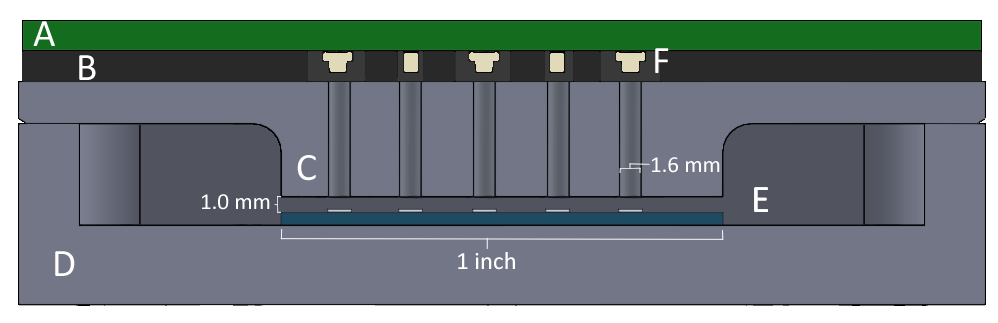}
        \caption{Upper: Printed circuit board (A) containing 30 LEDs (F), 3D printed lid spacer (B). Lower: Cross section of LED mapper testing setup showing the PCB (A), the spacer (B), the lid with collimator holes (C), the box base (D) and an illustrative detector wafer (E).}
        \label{fig:mapper}
    \end{figure}

    The LED mapper setup (Fig. \ref{fig:mapper}) consists of a printed circuit board (PCB), a spacer, and a collimating lid. 
    The PCB (Fig. \ref{fig:mapper}-A) contains 30 LEDs\footnote{LTST-C193TBKT-5A Lite-On Inc. } (Fig. \ref{fig:mapper}-F) connected to a row-column addressing scheme such that any combination of one input and one output wire activate a unique LED \cite{liu_supplementary_nodate}. Due to the close 2.2 mm packing of the MKIDs, it was not possible to have a one-to-one correspondence between pixels and LEDs. Pixels on the wafer are instead grouped into unit cells of three with their resonances intentionally designed to be far apart in frequency space to distinguish them. This method is already implemented to distinguish between the two orthogonal polarization MKIDs present in each pixel. Each unit cell thus contains six detectors, with two per pixel to account for polarization, across six distinct frequency banks as shown in Fig. \ref{fig:netanal}.
    The 3D-printed \footnote{Printed from ABS-ESD7 thermoplastic using fused deposition modeling} spacer (Fig. \ref{fig:mapper}-B) is intended to create a barrier between LEDs to prevent stray light from entering adjacent holes. Plastic was chosen to prevent risk of shorting the PCB circuit with a conductive material. 
    The aluminum collimating lid (Fig. \ref{fig:mapper}-C) contains 1.6 mm diameter holes that are aligned with the LEDs and pixel unit cells. The central portion of the lid extends to 1.0 mm  above the wafer to guide the light from the LEDs to the detectors while reducing stray light as much as possible.
    The pieces are all aligned using three dowel pins that are placed along the bottom and left side of the detector wafer (Fig. \ref{fig:mapper}-E) and connected from the testing box base (Fig. \ref{fig:mapper}-D) through the lid into the spacer.

\subsection{Testing Setup}
    The box is placed on the coldest stage of a cryostat cooled by a three-stage (4He-3He-3He) Helium sorption refrigerator, reaching a base temperature of 320 mK. 
    The LEDs are powered through wires running from the coldest stage to an Arduino microcontroller board at room temperature. The circuit is configured to allow for adjustable power input and LED flashing. 
    The MKIDs are readout using the RF-ICE readout system developed at McGill University \cite{rouble_rf-ice_2022}. 
    
\subsection{Testing Procedure}
 \begin{figure}
        \centering
        \includegraphics[width=0.9\linewidth]{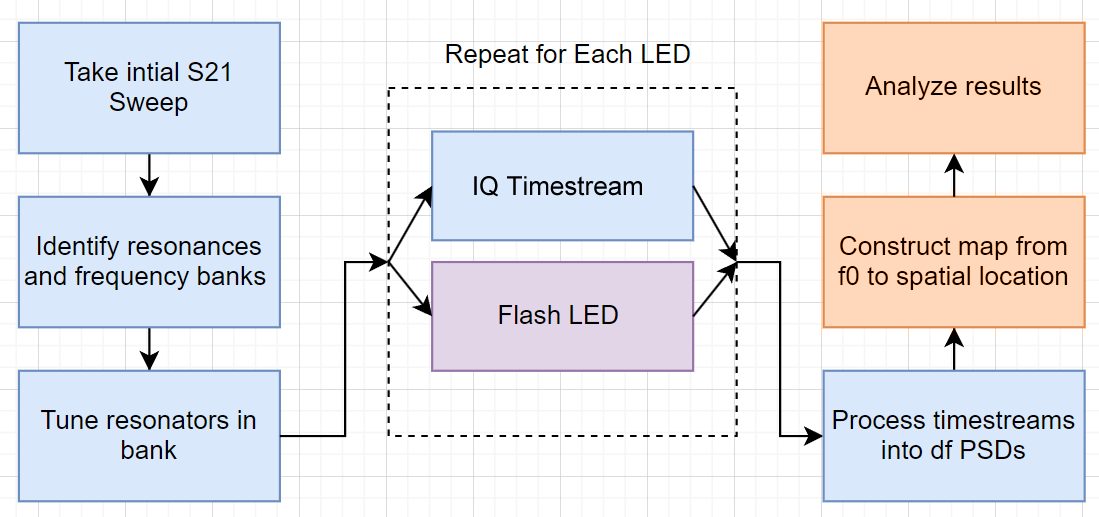}
        \caption{Flowchart of testing process. Blue indicates elements done through the RF-ICE system, purple indicates processes done through the LED mapper Arduino, orange indicates offline analysis processes.}
        \label{fig:flowchart}
    \end{figure}
    Fig. \ref{fig:flowchart} shows a schematic of the testing process, which we describe in more detail below.
    
    Testing begins by taking a sparse RF sweep across the frequency range of the readout bandwidth. From this sweep, resonances and their corresponding frequency banks are identified, first through an automatic finder, whose results are manually checked, with any missing resonances then added. A fine sweep with a width of 200 kHz is taken around each of the resonance positions for later computing the frequency shift, $df$. Bias tones are placed with sub-mHz precision on each resonance at the position that gives the highest response. 
    After the tones are set, a time stream is taken for thirty seconds. During this time, one LED flashes at a rate of 0.4 Hz with a $50\%$ duty cycle. This is repeated in 30 second increments until there is one time stream for each LED. The data is then processed from I/Q timestreams into $df$ timestreams from the fine sweep, and then to a power spectral density (PSD) of $df$. 

    From the $df$ PSDs, the frequency-pixel map is constructed by finding the resonance in each bank that has the highest response at the frequency of LED flashing. Matches are only accepted if the resonance response is higher than a threshold of 10$\sigma$ above the mean response of the rest of the resonators for that LED. Crosschecks are implemented to prevent mis-assignment due to variations in detector responsitivity. Only one resonance per frequency bank can be assigned to each LED.


\section{Preliminary Results}
    \begin{figure}
        \centering
        \includegraphics[width=0.8\linewidth]{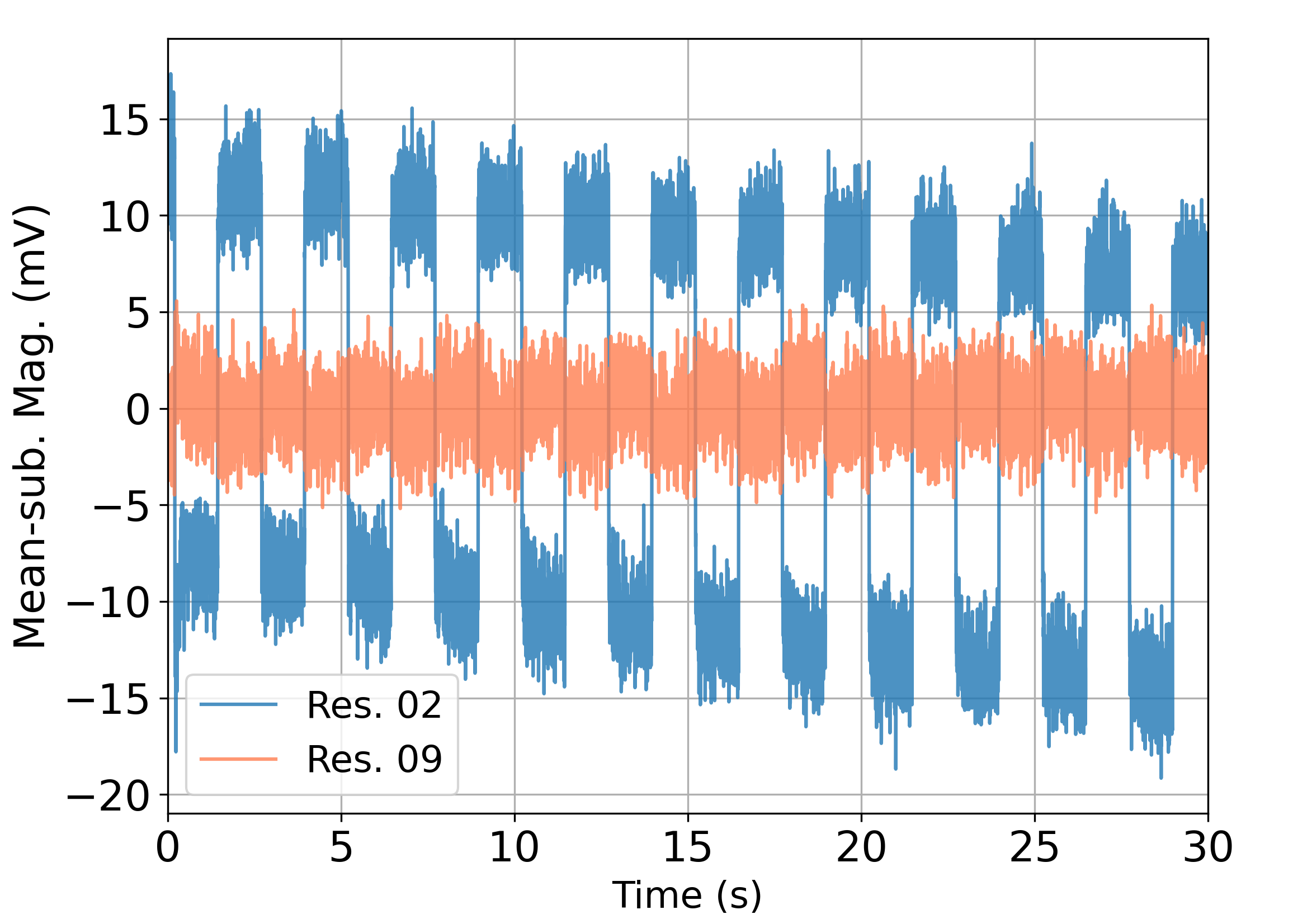}
        \caption{Timestream of MKID response during a LED flashing test. Blue shows an MKID under the flashing LED while orange shows an MKID elsewhere on the chip. A small amount of crosstalk between the two resonances is visible.}
        \label{fig:chop}
    \end{figure}
    LED mapping was performed on prototype 220 GHz SPT-3G+ MKIDs on an 1 in chip fabricated at the University of Chicago in the Pritzker Nanofabrication Facility. 
    Timestreams show a clear optical response to the LED flashing in the form of a chopped signal (Fig. \ref{fig:chop}). Resonators corresponding to MKIDs not underneath the active LED do not show the same response.

\begin{figure}
        \centering
        \includegraphics[width=0.8\linewidth]{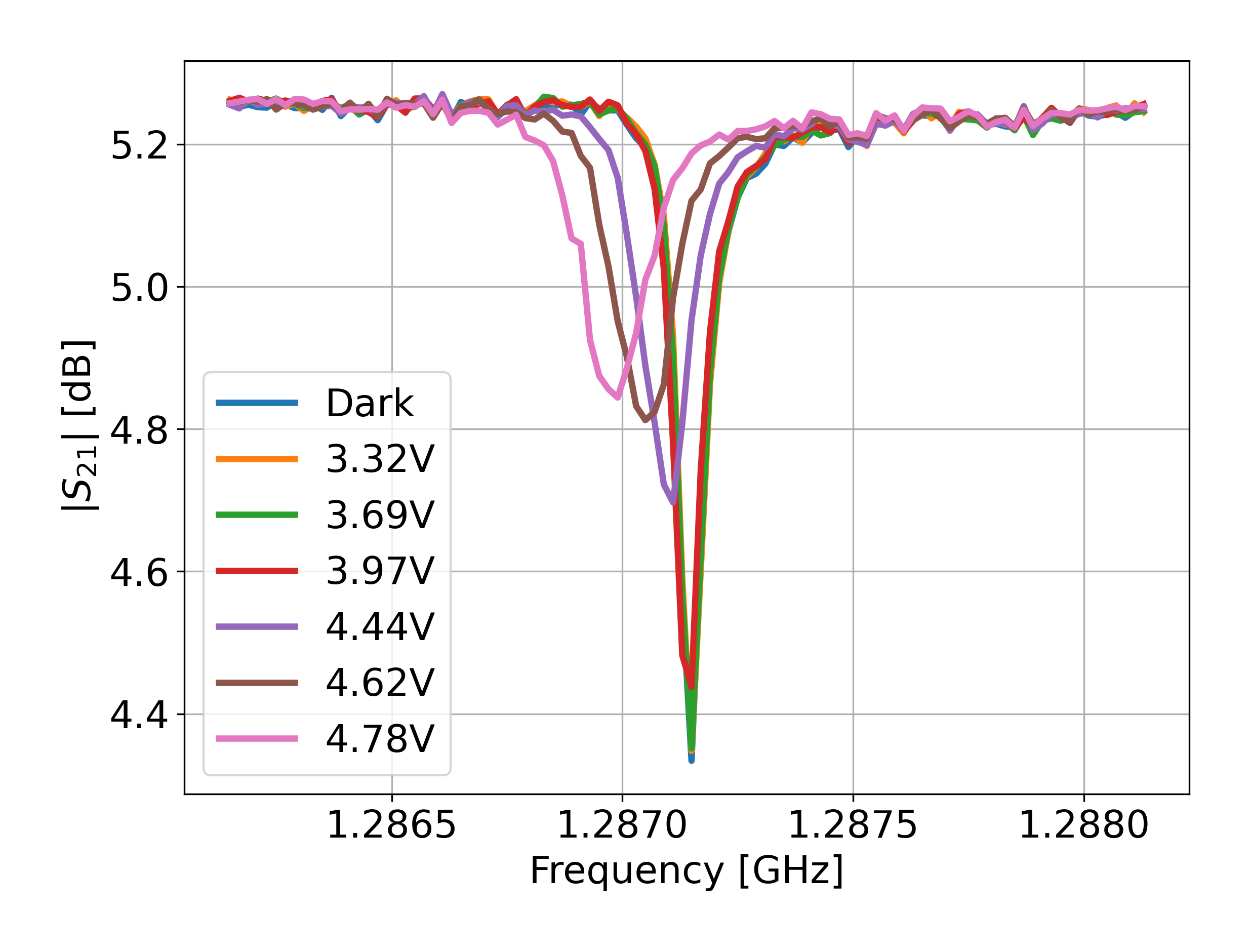}
        \caption{Change in the response of a single resonator while varying LED power levels, with the bias voltage indicated in the legend.}
        \label{fig:power_shift}
    \end{figure}

    Varying LED power levels cause varying response in the resonances. This is shown in Fig. \ref{fig:power_shift} as frequency shifts lower and depth becomes shallower with higher LED powers. 

\begin{figure}
        \centering
        \includegraphics[width=0.68\linewidth]{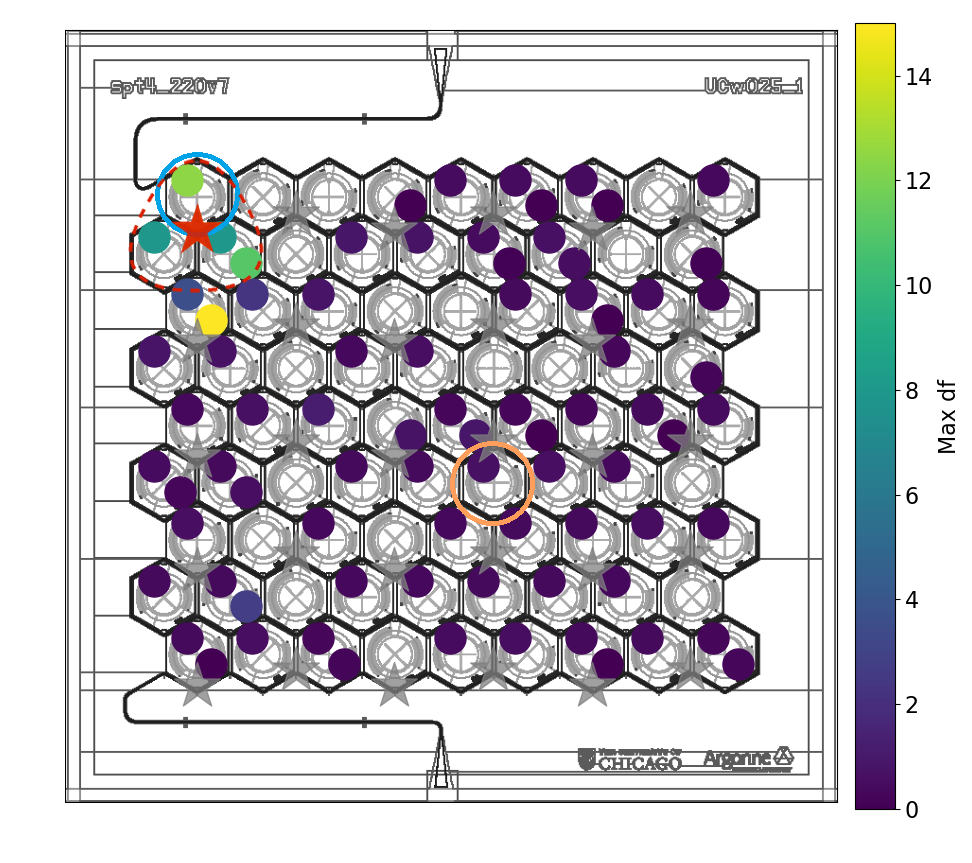}
        \caption{Map of resonance pixel locations, with the different polarization MKIDs in each pixel represented as solid offset circles. The hollow blue and orange circles correspond to Res. 02 and 09, respectively, in Fig. \ref{fig:chop}. Colorbar indicates maximum LED-normalized frequency shift, $df$, due to the LED indicated by the red star. The unit cell for that LED is given by the dotted line. The resonance with strong response outside of the unit cell is caused by stray light on a very responsive detector. The side lengths of the chip are 1 in.}
        \label{fig:colormap}
    \end{figure}

    From the frequency-to-pixel map generated by the mapping procedure, a plot is constructed showing the maximum normalized $df$ for each resonance for a given LED (Fig. \ref{fig:colormap}). This provides visual support of the map generation with all large responses occurring near the active LED.  

    
\section{Resonance Scatter}

\begin{figure}
        \centering
        \includegraphics[width=0.9\linewidth]{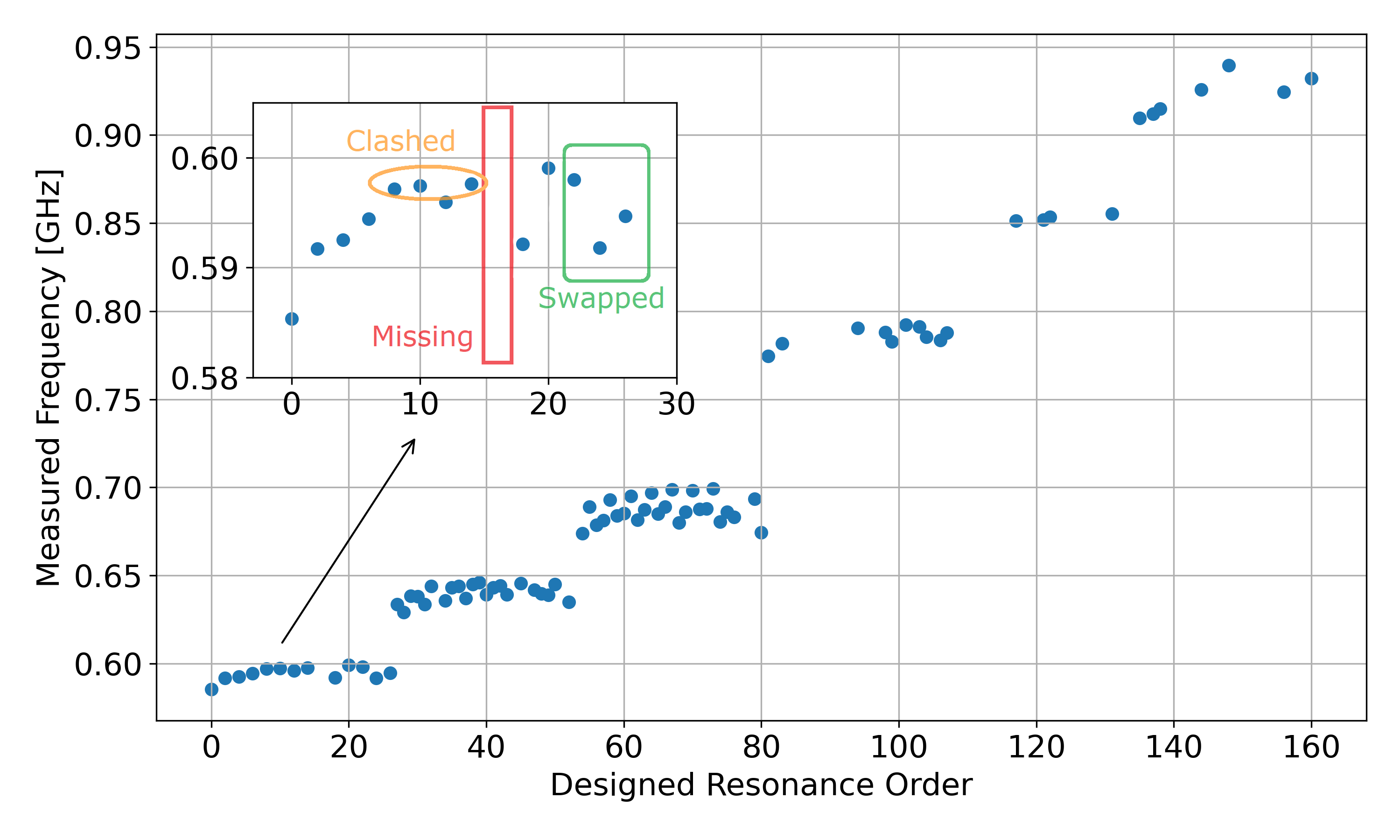}
        \caption{Plot of frequency of each detector determined by the LED mapper against the intended resonance order from the design mask. Inset shows zoom in highlighting examples of missing, clashed, and swapped resonances.}
        \label{fig:freq_v_order}
    \end{figure}

   The maps provide insight into the resonance scattering effects in our MKID arrays.
   One way to investigate this is through comparing the measured frequency of the MKID at each location in the array to the designed resonance frequency position of that MKID as computed via Sonnet simulation. In an ideal device, this would appear as six sets of linearly increasing points since the array is divided into six banks. Deviations from linearity indicate scatter and missing resonances. Fig. \ref{fig:freq_v_order} demonstrates the identification of clashed resonators with very similar frequencies, missing resonators, and resonators that have swapped positions.

\begin{figure}
        \centering
        \includegraphics[width=1\linewidth]{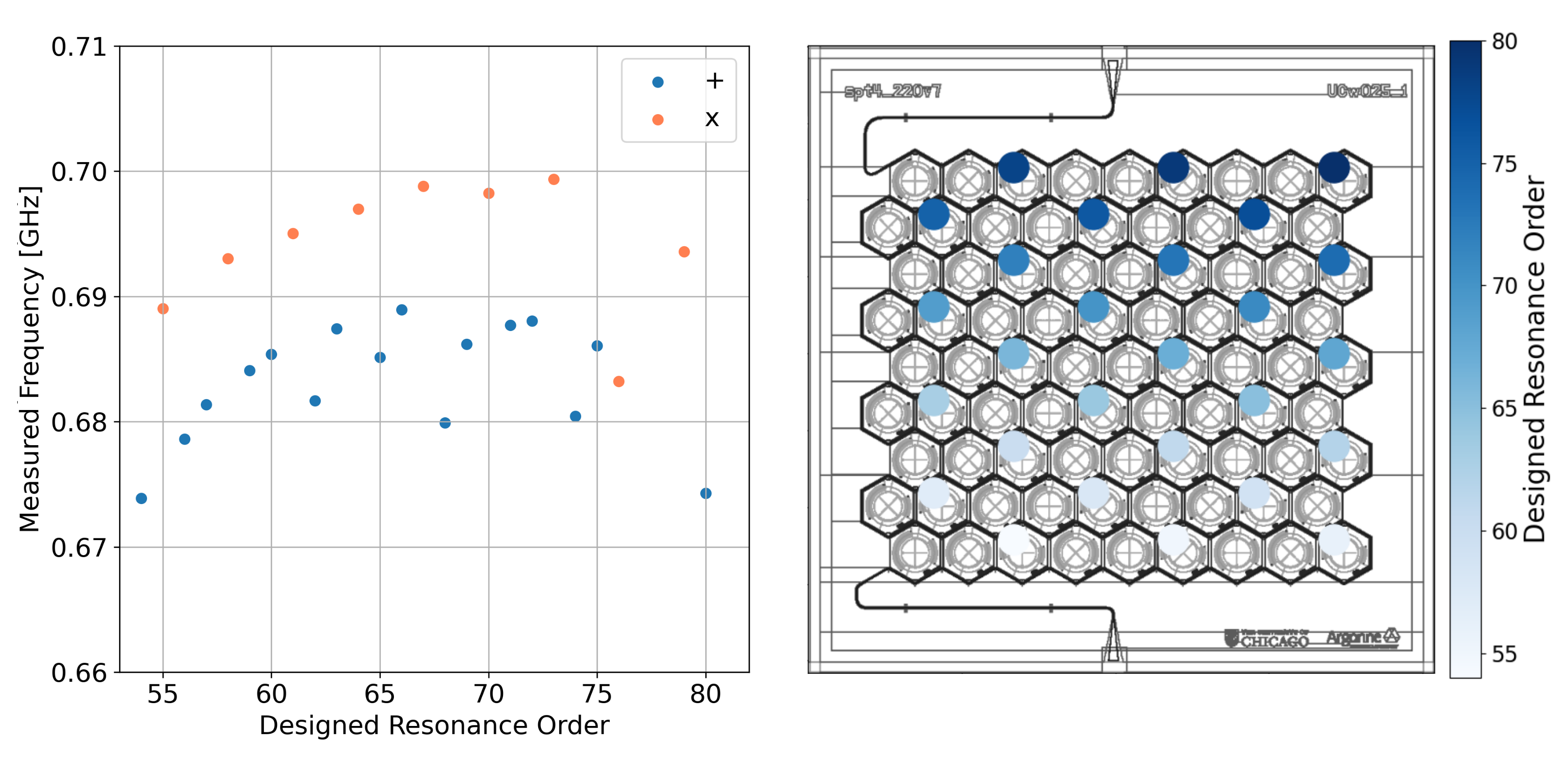}
        \caption{Left: Bank 3 detectors separated by ``$\times$'' and ``$+$'' type inductor orientations. Right: Designed location of bank 3 detectors with colormap of intended frequency order. The side lengths of the chip are 1 in for scale.}
        \label{fig:bank_3_plots}
    \end{figure}

    Each frequency bank exhibits two visible trends: downward curves from bank start to end, and a two-tiered structure in some of the banks. To identify the causes of these effects, we look more closely at the third bank of Fig. \ref{fig:freq_v_order}. The first effect can be explained by looking at the designed frequency of the detectors. The detectors are arranged to increase in frequency from low left to upper right, as shown in Fig. \ref{fig:bank_3_plots}. As the designed resonance order is correlated with pixel position, the downward curve indicates some type of spatial fabrication inconsistency across the device. This could be due to deposition or etch non-uniformities, but further study is needed to fully understand this effect.

    The cause of the ``tiered" structure is due to the differences between ``$\times$" and ``$+$" type inductor rotations. The $45^{\circ}$ inductor rotations are included for uniform sampling of the polarization states of electromagnetic radiation. Fig. \ref{fig:bank_3_plots} shows that detectors with the ``$\times$" inductor alignment have approximately 10 MHz higher frequency than those with the ``$+$" inductor alignment. Notably, this effect is missing from the first bank due to a mask error shorting the ``$\times$" orientation detectors in the first bank. 
    The cause of the discrepancy requires further study, but the main cause is suspected to be differing connecting line lengths between the inductor and IDC in the two variations. There may also be a contribution from a directional bias in the lithography process.

\section{Conclusion}
We have designed and tested a cryogenic LED frequency-to-spatial-pixel mapper for SPT-3G+ MKID prototype arrays. The MKID response to the LEDs is sufficient to identify and create maps of our devices, despite operating at 320 mK, as opposed to the nominal 100 mK operating temperature for these devices. This allows us to test and create maps of arrays in a cryostat cooled by a Helium-3 sorption refrigerator, while other testing of the devices requiring lower noise occurs using our lab's dilution refrigerator. The maps created have provided insights into causes of scatter affecting the SPT-3G+ prototype arrays which can be accounted for and counteracted in future versions of the array design. 

\section{Future Work}
\subsection{Resonance re-ordering with Capacitor Trimming}
\begin{figure}
        \centering
        \includegraphics[width=0.9\linewidth]{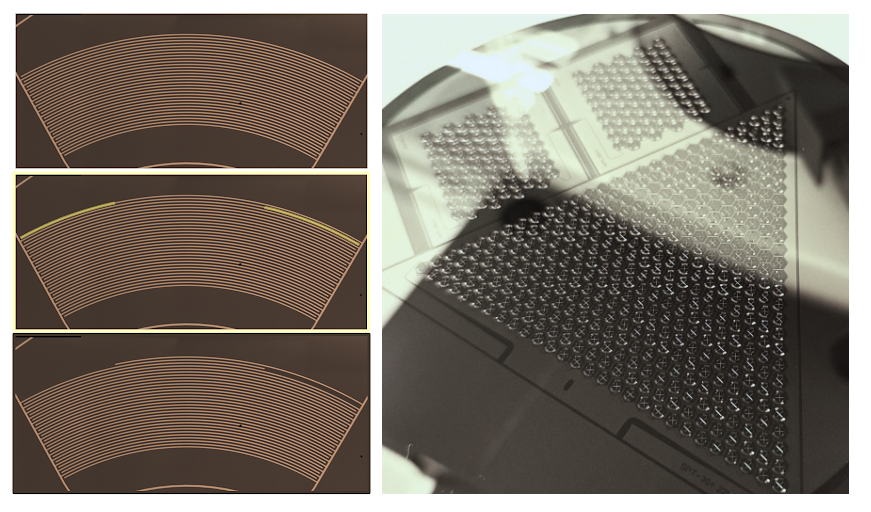}
        \caption{Left: A conceptual picture showing: (Top) The  IDC prior to trimming, (Middle) The IDC with sections targeted for trimming highlighted, and (Bottom) the IDC after trimming. (Credit: Kyra Fichman.)
        Right: SPT-3G+ 220 GHz dark triangular sub-module and two 1 in $\times$ 1 in prototype arrays on a 4 in wafer.}
        \label{fig:future_work}
    \end{figure}

    To achieve the very precise 566 kHz spacing and remove resonator clashes of existing arrays, it may be necessary to be able to adjust the frequency of each MKID after the initial fabrication. This can be done through trimming a section of the IDC in a second lithography process (Fig. \ref{fig:future_work}). This process is being pursued at the Pritzker Nanofabrication Facility at the University of Chicago.

\subsection{SPT-3G+ Sub-module Scale Mapper}
    The measurements described in this work have only been done on prototype 220 GHz SPT-3G+ arrays on 1 in chips with 160 MKIDs. Prototype deployment-scale 220 GHz dark triangular sub-modules for SPT-3G+ have been fabricated (Fig. \ref{fig:future_work}), and have so far attained a yield of $81\%$ after removing one half of all collisions. If all resonances in collisions could be preserved and reordered, the yield would increase to $>90\%$. We plan to build a LED mapper of the scale needed for a SPT-3G+ sub-module, and use this process of mapping and trimming, with a goal of achieving a $90\%$  yield for a design with 800 detectors per 500 MHz readout channel.

\section*{Acknowledgment}
This work and the South Pole Telescope program are supported by the National Science Foundation (NSF) through the awards OPP-1852617 and OPP-2408494. This work made use of the Pritzker Nanofabrication Facility of the Institute for Molecular Engineering at the University of Chicago, which receives support from Soft and Hybrid Nanotechnology Experimental (SHyNE) Resource (NSF ECCS-2025633), a node of the NSF’s National Nanotechnology Coordinated Infrastructure. 
Authors at Argonne National Laboratory acknowledge support by the U.S. Department of Energy (DOE), Office of Science, Office of High Energy Physics, under contract DE-AC02-06CH1137.
This work was supported by Fermilab under award LDRD2021-048 and by the National Science Foundation under award
AST-2108763
The McGill authors acknowledge funding from the Natural Sciences and Engineering Research Council of Canada and Canadian Institute for Advanced Research.

 
%

\bibliographystyle{IEEEtran}

\nocite{*}
\bibliography{ASC_Poster}


\vfill

\end{document}